# Memory-Aware Partitioning of Machine Learning Applications for Optimal Energy Use in Batteryless Systems

ANDRES GOMEZ, University of St.Gallen, Switzerland

ANDREAS TRETTER, ETH Zürich, Switzerland

PASCAL ALEXANDER HAGER, ETH Zürich, Switzerland

PRAVEENTH SANMUGARAJAH, ETH Zürich, Switzerland

LUCA BENINI, ETH Zürich, Switzerland and University of Bologna, Italy

LOTHAR THIELE, ETH Zürich, Switzerland

Sensing systems powered by energy harvesting have traditionally been designed to tolerate long periods without energy. As the Internet of Things (IoT) evolves towards a more transient and opportunistic execution paradigm, reducing energy storage costs will be key for its economic and ecologic viability. However, decreasing energy storage in harvesting systems introduces reliability issues. Transducers only produce intermittent energy at low voltage and current levels, making guaranteed task completion a challenge. Existing ad hoc methods overcome this by buffering enough energy either for single tasks, incurring large data-retention overheads, or for one full application cycle, requiring a large energy buffer. We present *Julienning*: an automated method for optimizing the total energy cost of batteryless applications. Using a custom specification model, developers can describe transient applications as a set of atomically executed kernels with explicit data dependencies. Our optimization flow can partition data- and energy-intensive applications into multiple execution cycles with bounded energy consumption. By leveraging interkernel data dependencies, these energy-bounded execution cycles minimize the number of system activations and nonvolatile data transfers, and thus the total energy overhead. We validate our methodology with two batteryless cameras running energy-intensive machine learning applications. Results demonstrate that compared to ad hoc solutions, our method can reduce the required energy storage by over 94% while only incurring a 0.12% energy overhead.

Authors' addresses: Andres Gomez, andres.gomez@unisg.ch, University of St.Gallen, Switzerland; Andreas Tretter, atretter@ethz.ch, ETH Zürich, Switzerland; Pascal Alexander Hager, phager@ethz.ch, ETH Zürich, Switzerland; Praveenth Sanmugarajah, sanmugap@student.ethz.ch, ETH Zürich, Switzerland; Luca Benini, benini@ethz.ch, ETH Zürich, Switzerland and University of Bologna, Italy; Lothar Thiele, thiele@ethz.ch, ETH Zürich, Switzerland.

X:2 Andres Gomez, Andreas Tretter, Pascal Alexander Hager, Praveenth Sanmugarajah, Luca Benini, and Lothar Thiele

# 1 INTRODUCTION

In recent years, the technology of batteryless systems [3, 18, 26] has evolved to tackle the power supply issues of traditional harvesting-based cyber-physical systems. Massive deployments requiring ultralong lifetimes with little to no maintenance are still out of reach due to energy storage technologies: rechargeable batteries suffer from high costs, hazardous materials, and a limited number of recharge cycles, typically less than $10^3$ [42]. Supercapacitors, another technology capable of storing large amounts of energy, have a higher number of recharge cycles but a prohibitively high self-discharge rate, and they would lower the system's energy efficiency [25]. Transient, or batteryless, sensing systems promise to overcome these limitations by requiring small energy storage capacities and operating opportunistically, whenever energy is available. When the energy requirement is small enough, batteryless systems can use standard energy storage technologies such as ceramic capacitors can offer virtually unlimited recharge cycles and very low leakage currents, thus enabling this vision of low-cost, ultralong system lifetimes.

**Harvesting-Based Systems.** The energy flow in all harvesting-based systems starts from ambient (primary) energy, moves through a transducer and power conditioning, and ends in an application circuit that does useful work. Ambient energy can be found in many different forms including light, heat, and vibration [5]. Transducer output depends not only on environmental conditions but also on other factors including its size (e.g., area, volume, or weight), I-V curve, and operating point. While sensing systems are typically designed to provide minimum service-level guarantees, energy sources rarely behave in a consistent and predictable manner. This is particularly true in photovoltaics, where light levels can change drastically in less than one second. Consequently, energy produced by transducers must be stored and conditioned before it can be used effectively [36]. How much energy a system can generate and store depends on many interrelated factors which can radically change the resulting behavior [40]. There are different ways to cope with harvesting variability, including service adaptation [31], power subsystem capacity planning [7], robustness analysis [43], and multi-modal energy harvesting [47].

Figure 1 shows a sample trace of a batteryless system executing a single iteration of a sense-process-transmit application over the course of an hour. The system ideally accumulates harvested energy and partitions the application over evenly sized energy bursts. The input power trace, based on real indoor photovoltaic measurements [41], provides a generous amount of harvested energy, almost 700 mJ, during one hour. But it also has a significant energy gap after 15 minutes, when a batteryless system would discharge completely. Naturally, the sensing system would execute reliably if it had a battery or supercapacitor, and that could even reduce the application's energy overheads since intermediate data would not need to be stored in nonvolatile memory. However, these features come at a significant price, since power use for even highly optimized deep-sleep states is in the $\mu$W range for commercially available systems [2]. Idle power is unavoidable in battery- or supercapacitor-based systems, and invariably leads to higher component and maintenance costs as well as energy losses due to self-discharge, idle currents, and converter inefficiencies. Batteryless systems, conversely, consume no power when fully discharged. This alone can lead to significant energy savings for batteryless systems deployed over many years. Furthermore, photovoltaics-based batteryless systems provide users with greater agency over when the sensor operates. Since visible light can be easily blocked by small layers of non-transparent materials, a user can cover the photovoltaic cell with a shutter and be assured it is physically impossible for the device to operate [15].

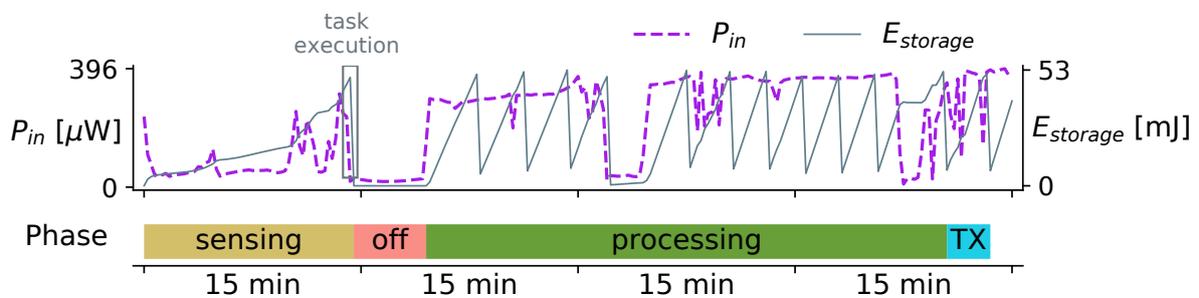

Fig. 1. Sense-process-transmit applications can be partitioned to run in an efficient, duty-cycled manner on batteryless systems. Environmental energy is highly variable, so batteryless systems can completely discharge at any point in time. To cope with this, they store intermediate data in nonvolatile memory.

**Batteryless Sensing Systems.** Batteryless sensing systems, also called transient systems, acquire, store, and transmit environmental data using environmental energy. However, they are supplied by volatile energy sources that can, at most, directly power the system for a very limited amount of time. During this time, the energy-harvesting rate might not be high enough to complete even one atomic task execution. We define an atomic task as a set of instructions whose functionality depends on its timely and uninterrupted execution. Common examples of these include reading a sensor, transmitting a wireless packet, or transferring a block of data from a volatile to a nonvolatile domain. Consequently, batteryless sensing systems must buffer at least the amount of energy needed to bridge the environmental power deficit and guarantee the completion of a single atomic task.

Low-power cyber-physical systems have components such as microcontrollers, nonvolatile memories, and external peripherals such as sensors and transceivers. Microcontrollers usually have a wide operating voltage range, but on-chip converters operate most efficiently at lower supply voltages [17]. External peripherals such as sensors and radios can have substantially different voltage requirements, but system designers usually avoid multiple voltage domains to reduce converter losses and simply choose the minimum voltage required to supply the entire system. In order to design a flexible platform that can efficiently harvest energy from different sources, it is necessary to decouple the source and load voltages, allowing each to operate at its respective optimal operating point.

Recent work has identified two approaches for batteryless operation: atomic and state-retentive execution. Atomic execution of tasks [18, 19, 32] relies on having enough storage capacity to complete either full applications or individual tasks without preemption. In [18], it was shown how dynamically adjusting the supply voltage to the optimal value *per task* can reduce the total application energy. The sense-and-process application was manually split into two separate bursts without any state-retention overhead. This is because the MSP430FR [44] has high-endurance nonvolatile memory for data is available on-chip, meaning that application data is already nonvolatile by default. For architectures without this onchip nonvolatile memory, data will need to be transferred between the (on-chip) volatile memory and the (off-chip) nonvolatile memory (NVM). Data-intensive applications will thus require optimized data transfers for energy-efficient execution. If a system's energy storage capacity cannot guarantee atomic execution of the application, then state-retentive execution becomes necessary. Unlike atomic execution, state-retentive execution needs to account for preemption, typically due to a power-critical interrupt. These systems [3, 26, 38] can be very efficient when the environment can sustain computation since it can avoid checkpointing. However, applications with power-hungry peripherals like cameras or radios might not be able to complete if their energy requirements are not met.

Depending on their tasks, sensing applications can have very different energy requirements. Tasks that use hardware peripherals such as sensors and transceivers cannot be interrupted before completion. These atomically executed tasks thus impose a minimum energy storage capacity required for correct functionality. Other tasks (e.g., data processing) can be interrupted before completion, but data needs to be transferred between volatile and nonvolatile domains. Applications with both types of tasks can be difficult to optimize for multiple reasons. Reliable execution depends on the ability to determine energy-burst sizes, which depend on both task sizes and the amount of data that needs to be saved to and restored from NVM. Determining memory requirements of conventional code is very challenging, and conservative estimates would lead to large energy overheads. Our proposed optimization scheme, called *Julienning*, is based on a data-flow specification model wherein programmers declare atomic kernels with explicit data dependencies. Our optimizer can then partition an arbitrarily long sense-process-transmit application into individual activations, or bursts, with deterministic energy bounds and minimal energy overheads. In doing so, atomic tasks can have the energy guarantees necessary for functional correctness and nonatomic tasks can be efficiently executed in accordance with those energy bounds.

Our proposed execution method for interruptible processing tasks leverages our data-flow specification model wherein small atomic kernels are used to compose large and complex processing tasks. *Julienning* groups together these smaller kernels into appropriately sized bursts, taking into account the overhead from transferring the necessary input and output data between volatile and nonvolatile memory. Since all kernels are developed with explicit data dependencies, *Julienning* can minimize this data transfer for any combination of kernels within a single burst. Our optimization flow is therefore capable of partitioning complex sense-process-transmit applications in a functionally correct and energy-efficient manner for a wide range of energy storage capacities. Furthermore, the tools we present are able to determine the minimum required energy storage capacity for reliable execution, as well as the minimum application energy for a given energy storage capacity. Implementing memory-intensive applications in batteryless sensing systems is a complex task that involves optimizing many parameters. *Julienning* greatly simplifies this implementation with an easy-to-use data-flow specification model that guarantees atomic kernel execution, minimizes energy overheads, and optimally partitions full applications into a batteryless execution model.

**Our Contributions.** Complex applications have two main hurdles for efficient transient (batteryless) execution. First, large and energy-hungry tasks require a very large energy storage capacity for atomic execution of the entire application. In reality, sensing applications consisting of multiple tasks only require task-based atomicity for functionally correct execution. Second, batteryless execution of data-intensive applications raises specific challenges since preempting a task due to energy unavailability incurs a large penalty from transferring data from the volatile to the nonvolatile domain. We use the head-counting applications first introduced in [16] as a sample batteryless sense-process-transmit application to demonstrate how our design tools can efficiently execute an energy- and data-intensive application. To this end, two head-counting systems were designed: one based on a normal vision sensor, the other based on a thermal (infrared) sensor. To detect the number of heads in each type of image, we use a trained convolutional neural network (CNN) [16], apply *Julienning* and execute the application in a low-power Cortex M4 microcontroller (LPC54102). By varying the storage capacity bound, we identify the Pareto front of both the thermal- and vision-based systems and their total energy overhead.

The main contributions presented in this work are summarized as follows:

- The *Julienning* optimization flow that partitions a large sequential application into bursts with bounded energy size and minimum energy overhead.

- Design and implementation of two transient camera systems using thermal and vision sensors.

- Trained CNNs with a small memory footprint. The head-counting application was implemented using a data-flow specification model with explicit data dependencies for batteryless execution.

- Experimental evaluation of both head-counting applications and their batteryless execution using the minimum feasible storage capacity.

- Design-space exploration using *Julienning* over a wide storage capacity range.

**Roadmap** The remainder of this work is structured as follows: In Section 2, we summarize current approaches to preempting batteryless applications. Section 3 introduces our general approach to specifying atomic kernels with explicit data dependencies. Section 4 presents the *Julienning* optimization algorithm. Section 5 introduces two machine learning applications and the batteryless platform they are implemented on. These are used to evaluate our proposed optimization flow in Section 6. Finally, we summarize our work in Section 7.

## 2 RELATED WORK

The intermittent power produced by transducers and its impact on the reliable execution of sensing applications has been studied in recent years. When the energy storage device is treated as a design variable, it can be sized to guarantee non-preemptive execution of a single sense-process-transmit cycle. While this approach avoids any runtime overheads, it typically requires a large energy storage device. If neither the energy source nor the energy storage can guarantee one full application cycle, then the application will inevitably be preempted, thus requiring special techniques to guarantee progress and functional correctness. Below, we discuss two general approaches which have been developed in prior work. If an application contains tasks with atomic execution requirements (e.g., sensing and transmitting), then sufficient energy storage must be provided for these tasks to execute reliably under adverse harvesting conditions. Applications without task-atomicity requirements can rely on mechanisms to detect power-critical situations and then consistently transfer data from volatile to nonvolatile memory before powering down. These software-based mechanisms focus on developing robust code that can withstand constant resets and still have consistent program progress.

### 2.1 Energy Storage-based Program Progress Guarantees

When information is short-lived, like in sense-and-send applications, a correctly sized storage device can guarantee the uninterrupted execution of a single application cycle, independent of intermittent power from transducers. If minimizing state-retention overheads is the only objective, this solution would be optimal since the data's lifetime starts and ends within one activation cycle. Examples of such systems include cameras that wirelessly transmit pictures [32], wearable cameras that can estimate a user's walking speed [19], and ambient sensors that transmit measurements via Bluetooth Low Energy (BLE) [30] and LoRa [29]. Sizing energy storage for uninterrupted application cycles is not a scalable approach as storage size increases linearly with application requirements. Other works [11] have proposed a reconfigurable energy storage architecture, which can adjust to dynamic application energy demands. By contrast, our *Julienning* approach uses the energy storage capacity as an optimization constraint. This allows a designer to explore a wide range of energy storage bounds to find the minimum capacitance necessary for guaranteeing atomic tasks rather than entire application cycles. In this way, the capacitance can be greatly reduced. Our optimization model minimizes the energy overhead given the energy storage bound.

## 2.2 Software-based Program Progress Guarantees

When application's execution energy requirements exceed a system's storage capacity, software execution requires specialized methods for guaranteeing consistent program progress. Backing up data in nonvolatile memory is a common support mechanism for data consistency in batteryless applications, and much work has gone into optimizing this process. Architectural support for checkpointing includes specialized techniques, such as scan-chain-based methods [21], for data transfer between volatile flip-flops and shadow nonvolatile flip-flops. Furthermore, advanced data-tracking techniques restrict data transfer to only those registers which changed after the previous backup [24]. FRAM-enabled designs have been shown to have costs of 3.44 pJ per bit [35] and <400 ns wake-up time [27] using 130 nm technology, while a 65 nm ReRAM-based design [28] has 20 ns restore time, operating at 100 MHz.

Software-based methods are based on an external signal which warns the system of an imminent undervoltage condition, triggering a data transfer from the volatile to the nonvolatile domain. Software libraries have been developed to intermittently execute processing tasks by automatically saving and restoring volatile data based on voltage thresholds [3, 26]. In [38], support for reconfiguring external peripherals was added. These approaches have been demonstrated to work on 16-bit MSP430 microcontrollers with on-chip FRAM. Automatic checkpointing techniques have also been demonstrated on flash-based systems [37] and 32-bit Cortex M3 systems [4] along with energy-aware optimizations [6]. Additional state-retention policies have also been proposed to exploit the different properties of both FRAM and flash to find the most efficient policy and platform [46], Besides checkpointing, applications have also been decomposed into tasks that are then executed individually [18, 20, 22]. These systems require manually transferring data to nonvolatile memory. While state-retentive systems excel at minimizing the required energy storage for executing arbitrarily long processing tasks, they have a fundamental limitation from their small storage: large atomic tasks using power-hungry sensors or transceivers are simply not supported.

Specialized languages and runtimes have been developed for batteryless systems. The Dewdrop [8] runtime, developed for flash-based MSP430s, is able to dynamically adjust the system's wake-up voltage to find the lowest value for reliable task execution. With Chain [9], developers can specify applications as static task graphs with statically multiversioned channels and restricted access to volatile- and nonvolatile-memory domains. Chain is able to reduce checkpointing costs, with guaranteed consistency, by marshaling data and allocating multiple copies of data in nonvolatile memory. CleanCut [10] can automatically decompose applications into tasks by placing task boundaries such that they can be executed with a given energy storage. CleanCut works by analyzing the application's control-flow graph and using a statistical energy model to avoid nonterminating path bugs. The Mayfly [23] language and runtime, developed for MSP430s with on-chip FRAM, allows developers to declare tasks with time-dependent data flows. In this way, the runtime can decide whether or not to execute tasks based on the age of sensed data.

All of these programming models assume an inherently unreliable hardware layer that can reset the system at any point. In contrast, we build on top of an energy management unit (EMU), which can guarantee a specified amount of energy, regardless of variability in a transducer's voltage and current. This opens the door for a programming model with guaranteed program progress with controlled, consistent preemption at predefined program points. This allows us to run energy-hungry tasks, even if their power requirements are beyond the limits of the source.

Listing 1. Sample definition of a sense-process-transmit application. Kernels are executed atomically and have explicit dependencies marked with "in" and "out". Metakernels can call multiple atomic kernels and be partitioned into bursts. In complex machine learning applications, long processing tasks must be declared as metakernels.

```c
#define Dx 80
#define Dy 60

kernel sense( out uint8_t img[Dx][Dy] ) {
  /* Kernel must enable/disable the peripheral it uses */
  camera_enable();
  AcquireImage(img);
  camera_disable();
}

kernel process( in uint8_t img[Dx][Dy], out uint8_t headCount[1] ) {
  /* Picture is processed, result saved to 'headCount' */
  headCount[0] = Detect(img);
}

kernel transmit( in uint8_t headCount[1] ) {
  /* Kernel must enable/disable the peripheral it uses */
  radio_enable();
  BLE_send(headCount[0]);
  radio_disable();
}

metakernel main() {
  uint8_t img[Dx][Dy];
  uint8_t headCount[1];

  /* Acquire picture and store it in 'img' */
  sense(img);
  /* Process picture and save result in 'headCount' */
  process(img,headCount);
  /* Transmit result via BLE  */
  transmit(headCount);
}
```

Furthermore, our programming model can automatically minimize the total application energy using a transducer-independent, energy-burst execution model.

## 3 THE *JULIENNING* APPROACH

This section presents an overview of the *Julienning* approach. After showing the application specification scheme and the steps involved in transforming specifications into platform-specific C code, we discuss the different optimization goals and levers connected with this transformation.

We use the *Ladybirds* application specification model [45] to obtain analyzable application code. Ladybirds was specifically designed for efficient data exchange for parallel applications in multicore architectures. However, its key approach of making data dependencies explicit is equally expedient for addressing the data backup and restoration requirements of transient systems. Applications are specified in the programming language *Ladybirds C*, which essentially extends the well-known C language with *kernels* and *metakernels*. Essentially, kernels are conventional C functions with explicitly specified inputs and outputs of fixed sizes. This implies that they have no side effects (e.g., they must not modify global variables). Metakernels are a special type of kernel: while normal kernels are used to implement specific functionality using arbitrary C code, metakernels only call kernels or other metakernels, passing arrays or subarrays as parameters in a MATLAB-like fashion. In other words, they interconnect kernels to build applications.

Listing 1 shows an example of a simple sense-process-transmit application specified in Ladybirds C. It contains three kernels with explicit data dependencies and one metakernel joining them. kernel parameters have been marked either only as "in" or "out" to indicate memory that is only read or written, respectively. Ladybirds also supports "inout" which would allow data to be both read and written. For simplicity, we show here a simple *process* kernel, which takes a read-only *img* matrix containing the picture, and writes the results from the head-detection algorithm to the *headCount* memory location. In a complex machine-learning application, there are multiple kernels defining the convolutional filters for a small image patch, and many intermediate results before the final headcount can be summed up.

Because of their simple structure, metakernels can be analyzed statically and their call hierarchy is fully flattened out (one could call it full inlining), so that a Ladybirds application can be described as a sequence of calls to different kernels. Each such kernel call is called a *task*. By performing an array-based static single assignment analysis [39], the application can be brought into a form where each task produces a fixed number of data *packets* of fixed sizes and requires a fixed number of given packets from preceding tasks. In the simple example from Listing 1, this would result in one task for each kernel, but if one kernel is called multiple times, multiple tasks will be instantiated.

When executing a Ladybirds application on a transient system, one could interrupt execution inside one of the tasks. However, since these tasks consist of arbitrary code, this would entail all the intricate state-saving problems encountered in previous work (cf. Section 2). *Between two tasks*, on the other hand, interrupting the program is easy because the state of the application is given by the explicit data dependencies of the tasks and only newly produced packets that are known to be required later need to be stored to and restored from nonvolatile memory. Using an EMU allows us the controlled and guaranteed execution of *bursts*, in which all required packets are loaded from NVM, one or more tasks are executed, and newly produced packets are stored back to NVM.

Algorithm 1 illustrates this burst-based application execution that we target using *Julienning*. As soon as sufficient energy has been stored in the capacitor, the system receives an external interrupt from the EMU, boots up, and looks in the NVM to find which task(s) are next in the sequence to execute. Given this sequence of tasks, the system then loads from NVM all required inputs, executes them, and writes all obtained outputs back to the NVM to be used later on. As soon as this is completed, the system updates the corresponding status information in the NVM and shuts down.

The question is now which tasks to execute together before preempting. We refer to this problem as *partitioning* the application into bursts. This step has large effects on the final program; in particular, two competing goals exist:

- Minimizing the total overhead and thus the overall energy needed to execute the application.

- Respecting the system's energy storage capacity bound.

If the energy storage capacity is unbounded, it is easy to see that minimal overheads are incurred given that the application will not be interrupted. This would mean the entire application is executed atomically. While this solution has a low energy overhead, it also requires a large energy storage capacity. Consequently, its charging time will be slow, since the system will need to slowly accumulate the energy required for the entire application before it can even start executing. If the storage capacity can be adjusted between the energy levels required for the largest atomic task and for the total application, then a trade-off will emerge. The overhead from preemption will increase slightly, but the required energy storage capacity and the corresponding charging time

**ALGORITHM 1:** Burst-based application execution

**while** *true* **do**
    Wait for sufficient energy availability;
    Start up system;
    Retrieve from NVM the index of the current burst to be executed;
    Initiate DMA transfer of required input data from NVM;
    **foreach** *task in current burst* **do**
        Wait for DMA availability of all input data required by the task;
        Execute the task;
        Initiate DMA transfer to NVM of all task outputs required in later bursts;
    **end**
    Wait for all DMA transfers to complete;
    Increment current burst index in NVM;
    Shut down system;
**end**

will decrease. Exploring this design space is highly desirable, but partitioning an application given a storage capacity is a difficult problem to solve. This is particularly challenging with machine learning applications which have large processing and memory requirements, and can easily incur large overheads when the application is interrupted.

One approach could be fixed partitioning, which takes a constant number of tasks per burst. This is fairly simple to implement manually, but can lead to great inefficiencies. First, it is not easy to track the data dependencies, so more data would be loaded and stored to nonvolatile memory than would actually be needed. Second, it is highly unlikely that all tasks will consume the same amount of energy, meaning that some energy bursts will not fully utilize the energy budget given by the energy storage capacity. This is inefficient because it is possible to avoid nonvolatile-memory transfers for a data dependency between two tasks if they are executed together.

From these considerations, we can derive two additional desirable properties for good partitioning. First, all bursts should have similar energy costs. Second, a burst boundary (i.e., preemption) should only occur where the data dependencies between the tasks before and after it are low in volume. Since these good partitions are not always easy to find amongst the $2^{n-1}$ possibilities for an application with $n$ tasks, we need an automated way to calculate optimal partitions that can make the most of a bounded storage capacity and also minimize energy overhead.

We will propose a burst energy model for calculating these costs based on precharacterization data. For the state retention, we can determine the amount of data that needs to be transferred and use a linear model to determine the energy cost. The system's energy storage capacity is an independent design parameter to be chosen by the developer. This will have a clear impact on how, if at all, the application will be partitioned. Using the application model and energy characterization, we can then find the optimal burst partitioning such that the burst size respects the energy bound and the total overhead is minimized. As we will also show, we can calculate the minimum energy storage that is needed to execute an application on a transient system. All this is part of our *Julienning* approach and will be explained in the next section.

## 4 OPTIMAL PARTITIONING

In this section, we describe the type of system we consider, and formalize its properties. With these prerequisites, we give a formal description of the goal of *Julienning* in the form of an optimization problem, that is then solved in two steps: In order to minimize the overall energy consumption, we first have to calculate how much energy a burst actually consumes. Once we have this value for all possible bursts, we find the optimal partitioning. Finally, we show that we can also calculate the minimum energy storage capacity necessary to execute the application.

### 4.1 System Model and Problem Description

We consider a system that repeatedly executes the same application and model it as follows.

*Application.* The program to be executed is a sequence of $n_t$ tasks $t_1, \ldots, t_{n_t}$, which are executed in that order. Each task $t_i$ reads and writes, respectively, a set of *data packets* $P_i^r \subseteq P$ and $P_i^w \subseteq P$, with $P = \bigcup P_i^w$ being the set of all packets in the application. Each packet $p \in P$ is contained in exactly one $P_i^w$ but can belong to one or more $P_i^r$.

*Energy costs.* The system incurs the following energy costs, which are assumed to be known.

- Each task $t_i$ consumes an amount $E_{task,i}$ of energy for its execution. These values can be obtained with measurements or static worst-case analyzers.

- Loading and storing of a packet $p$ from and to the nonvolatile memory requires energy amounts of $E_r(p)$ and $E_w(p)$, respectively. These energies are hardware-specific and mostly depend on the packet size. Typically, one would expect a linear dependency with a constant initialization offset. While these costs differ between the cases when DMA transfers can be carried out in the background and when the system must wait for the transfer to finish, we always assume the latter case for the sake of simplicity. This conservative approach avoids the uncertainty and the intricacies involved in predicting task execution times and thus the amount of data-transfer concurrency.

- The start-up energy $E_s$ specifies the fixed costs for system initialization on each burst. We also include the costs for reading and writing the current burst index.

These specifications define the background for the *Julienning* optimization, which can be expressed as the following optimization problem:

> Given a maximum amount $Q_{\max}$ of energy that can be stored in the capacitor, partition the tasks into a set of bursts such that (i) no burst consumes a higher amount of energy than $Q_{\max}$ and (ii) the total amount of energy consumed per application execution is minimized.

The remainder of this section will show how this problem can be solved. We first need to know how much energy any arbitrary burst would consume.

### 4.2 Calculating Burst Energy Costs

In the following, we discuss how to calculate $E\langle i, j\rangle$, the energy required to execute a burst that contains tasks $t_i, \ldots, t_j$. If a burst contains only one task, i.e., $i = j$, then executing the burst consists of starting up the processor, loading the task's input packets, executing the task, and storing its output packets to nonvolatile memory. The burst energy can thus be calculated as

$$E\langle i, i\rangle = E_s + \sum_{p \in P_i^r} E_r(p) + E_{task,i} + \sum_{p \in P_i^w} E_w(p).$$

If a second task is added to the burst, the burst energy can be computed analogously, but a number of additional effects need to be considered:

- An input packet of the second task could also be required by the first task. In that case, the packet would not have to be loaded from nonvolatile memory twice.
- An input packet of the second task could have been produced by the first task. In that case as well, the packet would already be in volatile memory.
- An output packet of the first task could be used solely by the second task. In that case, the packet would not have to be saved to nonvolatile memory.

Similar considerations hold for tasks consisting of multiple bursts. To formalize these constraints, we define the *last use* $l_j(p)$ of a packet $p$ prior to an index $j$ as the highest index less than $j$ of a task that reads or writes $p$:

$$l_j(p) = \max \left\{ i < j \mid p \in P_i^r \vee p \in P_i^w \right\}.$$

This definition includes $l_\infty(p)$, the index of the last task in the application that reads or writes $p$. We can now express the sets of packets $P_k^r \langle i, j \rangle$ and $P_k^w \langle i, j \rangle$ that must be loaded and stored, respectively, for a task $t_k$, $i \leq k \leq j$, in a burst containing tasks $t_i, \ldots, t_j$:

$$P_k^r \langle i, j \rangle = \left\{ p \in P_k^r \mid l_k(p) < i \right\}, \qquad P_k^w \langle i, j \rangle = \left\{ p \in P_k^w \mid l_\infty(p) > j \right\}.$$

The burst energy can then be computed as

$$E \langle i, j \rangle = E_s + \sum_{k=i}^{j} \left( \sum_{p \in P_k^r \langle i,j \rangle} E_r(p) + E_{task,i} + \sum_{p \in P_k^w \langle i,j \rangle} E_w(p) \right).$$

These energy costs can now be calculated for every possible burst, providing the basis for the following optimization.

### 4.3 Optimal Partitioning

Once the energy costs for all possible bursts are known, we can select the optimal bursts to solve the task-partitioning problem, i.e., the question of how the tasks should be partitioned to bursts such that the overall application execution energy is minimized while respecting the upper energy bound $Q_{\max}$. However, there are $2^{n_t-1}$ possible partitions of a given sequential application into bursts. An *exhaustive search* for finding the optimal partition is therefore computationally unfeasible for arbitrarily long applications.

To address this problem, we first consider the case that each task is executed in its own burst. In this case, we have $n_t$ bursts and we can define $(n_t + 1)$ power-off states $s_0, \ldots, s_{n_t}$. In this definition, $s_0$ is the state before application execution has started, $s_1$ is the state after the burst containing $t_1$ has executed, and so on (see Figure 2). The system consumes energy $E \langle 1, 1 \rangle$ while transitioning from $s_0$ to $s_1$, $E \langle 2, 2 \rangle$ while transitioning from $s_1$ to $s_2$, and so forth. If we now take bursts with multiple tasks into consideration, the system can also get from $s_0$ directly to $s_2$ by executing a burst with $t_1$ and $t_2$, consuming the energy $E \langle 1, 2 \rangle$. In this way, we can construct a *state graph* connecting each state $s_i$ to each state $s_j$, $j > i$, and assign to each edge a cost equivalent to the corresponding burst energy.

This graph offers a new interpretation of the task-partitioning problem. Minimizing the total application execution energy can be interpreted as the problem of getting from $s_0$ to $s_{n_t}$ with the smallest cumulative cost. The constraint of keeping the burst energy below $Q_{\max}$ means that no edges with a cost larger than this upper bound may be considered. The task-partitioning problem, therefore, is equivalent to the problem of finding the shortest path from $s_0$ to $s_{n_t}$ on a state graph from which the edges with a cost higher than $Q_{\max}$ have been removed.

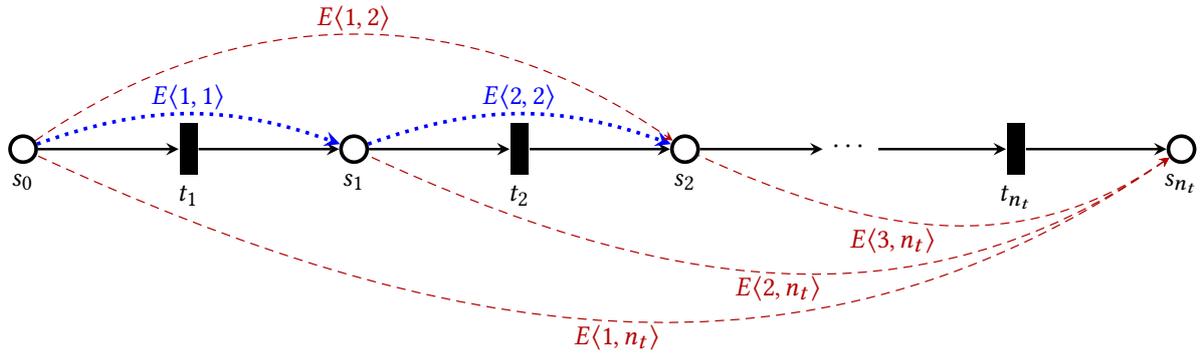

Fig. 2. Graphic illustration of the states $s_0, \ldots, s_{n_t}$ between the single-task bursts executing tasks $t_1, \ldots, t_{n_t}$. Burst energies for transitions between the states have been added. Dotted blue lines represent single-task bursts; dashed red lines represent multi-task bursts. This leads to the construction of a state graph.

Shortest path problems can be efficiently solved, for instance, using Dijkstra's algorithm [13], with a complexity of $O(n_t^2)$. The complexity of the proposed calculations is, in fact, dominated by the computation of the burst energies $E\langle i, j \rangle$: There are ½ $n_t$ ($n_t + 1$) possible burst energies to calculate. For each such burst, one needs to iterate over all tasks contained in it and check whether each packet accessed by each task needs to be transferred or not. These checks can be implemented efficiently (amortized $O(1)$) using lookup tables, which leads to an overall complexity of $O(n_t^3 \cdot |P|)$. While more complex than the mere shortest path calculation, the method finds the optimal solution to the task-partitioning problem in polynomial time. In practice, the number of computations can also be significantly reduced using algorithms that skip the calculation of a burst's energy cost if the mere execution costs of a burst containing a subset of its tasks already surpass $Q_{\max}$.

### 4.4 Storage Minimization

The method proposed in the last section can not only be used to minimize the overall application execution energy, but also to determine the smallest energy storage capacity $Q_{\min}$ necessary to execute a given application. Note that this energy is *not* equivalent to the highest burst energy of all single-task bursts, since adding tasks to a burst could save more data-transfer energy than the additional task execution energy costs. $Q_{\min}$ can be calculated by modifying the shortest path algorithm such that a path length is not calculated by adding all costs along the path, but instead by choosing the maximum single-edge cost along the path. Applying the modified algorithm to the complete state graph (with no connections removed) will yield $Q_{\min}$ as the shortest path cost. Another run of the original partitioning algorithm then returns the optimal partitioning for this maximum burst cost.

### 4.5 *Julienning* Summary

Figure 3 summarizes the main steps within *Julienning* approach. An application described in Ladybirds C is first transformed into a task graph with explicit data dependencies. This graph is then transformed into a state graph, where each state represents the application state before and after each sequential task is executed. In this state graph, each edge represents the tasks necessary to get from one state to another, along with their data dependencies. Once this graph has been pruned of edges that already exceed the energy bound imposed by $Q_{\max}$, the shortest path between $S_0$ and $S_N$ thus represents the application partitioning scheme with the lowest energy cost, subject to the constraints imposed by $Q_{\max}$. *Julienning* is capable of automatically performing these powerful transformations and optimizations. In particular, the problem of optimizing data loading and storing depending on the task dependencies would be intractable to do manually. The resulting partitioning scheme can then be easily loaded on a runtime environment responsible for the burst-based execution model.

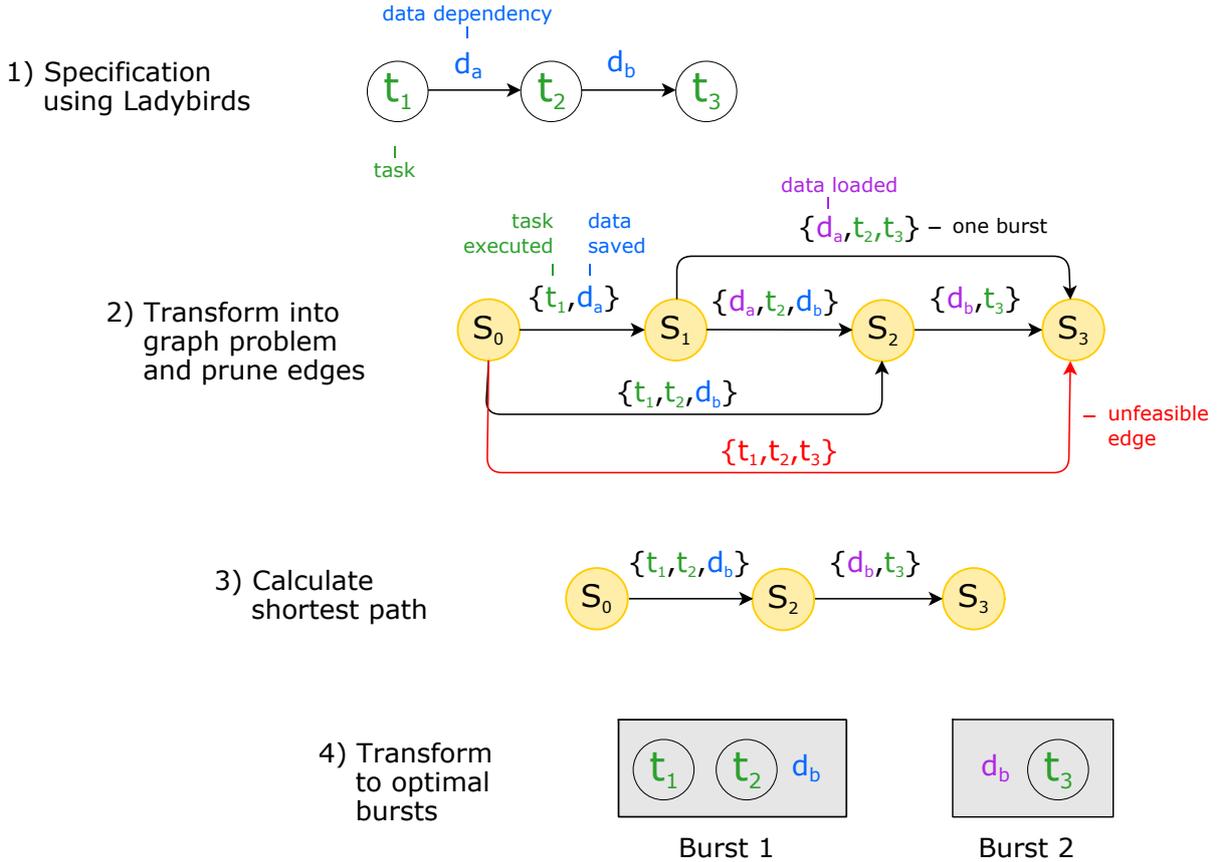

Fig. 3. Summary of the automated steps of the *Julienning* optimization flow. In this example, executing all tasks in one burst would require storage capacity larger than $Q_{\max}$, and executing tasks $t_1$ and $t_2$ together avoids the overhead of saving $d_a$ in nonvolatile memory.

## 5 MACHINE LEARNING APPLICATIONS ON BATTERYLESS SYSTEMS

To show the generality of batteryless systems, we demonstrate the efficient and reliable operation of a machine learning application with large energy and processing requirements. More specifically, we focus on a convolutional neural network (CNN) algorithm to estimate the number of people in a closed space. This is an important use case in public spaces such as airports and stations as well as in smaller environments such as offices and classrooms (e.g., to track the occupation of classrooms).

Though image processing applications have been implemented on batteryless systems before [18, 19], the energy and processing requirements of these applications have been relatively small, enough to process images at a maximum rate of several frames per second. In this work, we have adapted a CNN-based head-counting application for battery-based systems, first presented in [16], with significantly higher processing requirements. Almost 100 seconds of processing are required for the application to finish. We built two separate systems, one using a thermal (infrared) camera [14] and another using a traditional visible light camera [34]. After the CNN detects the number of heads in the image, the result is broadcast using Bluetooth Low Energy (BLE). The main focus of our work is partitioning machine learning applications for batteryless execution, not the detection accuracy of the different systems. To compare *Julienning* fairly, we thus scale down the visual image and use the same CNN architecture for both the visual and the thermal images. In this way, the only difference between the two versions is the energy required for the image acquisition itself.

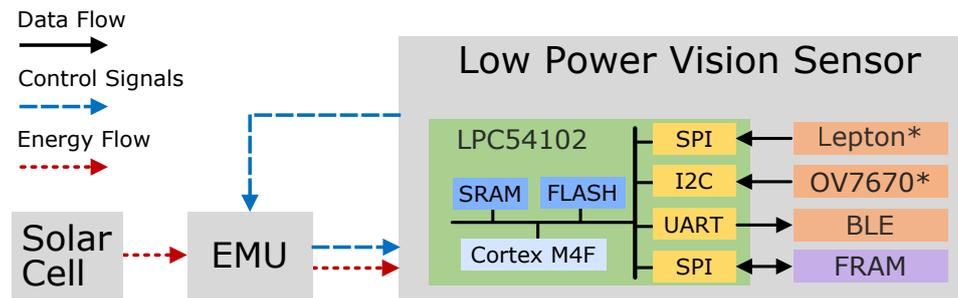

Fig. 4. Overview of the ultra-low-power people-recognition platform. One system uses a visual sensor (OV7670 [34]), while another uses an infrared sensor (Lepton [14]). The energy management unit (EMU) [?]GSMBT2016) buffers harvested energy and triggers prople-recognition platform when the energy buffer is full.

Batteryless systems can only store very limited amounts of energy, typically enough to power its largest atomic task [18]. With such a small energy storage element, a batteryless system can quickly recover from a full discharge and make efficient use of harvested energy. As shown in Figure 1, batteryless systems can also discharge completely in a short time whenever the environment provides too little energy. To be able to cope with power outages without data loss, batteryless systems require nonvolatile memory. As described in the previous section, our *Julienning* approach can minimize the overhead of transferring data between volatile and nonvolatile domains. It does so by first grouping together as many processing tasks as possible within the energy-burst bounds. It then restricts the loaded data to that required by those tasks. It also restricts the saved data to that required by future tasks. This overhead minimization is particularly important in machine learning applications since they have high memory and processing requirements. These features, combining atomic execution and optimized memory transfers, require low-power hardware support for energy-efficient operation. We now discuss the implementation details of our batteryless systems.

### 5.1 Embedded Implementation

Two standalone head-counting systems were developed with commercial off-the-shelf components. The vision-based version used the OV7670 [34] camera as sensor input, while the thermal-based version used the FLIR Lepton [14] sensor. The head-counting algorithm (same for both versions) was implemented on a LPC54102 microcontroller [33]. A BLE radio [1] was used to transmit the results of the head-counting process. Lastly, the nonvolatile-memory requirement was satisfied with an external FRAM memory [12]. This technology has very high endurance and energy efficiency, albeit at a higher component cost compared to traditional technologies such as flash.

In Figure 4, we show a simplified diagram of the full platform. The energy management unit (EMU) [18] contains an energy harvester with maximum power point tracking (MPPT) and a capacitor with serves as a energy buffer with a limited capacity. The EMU sends a wake-up trigger to the application circuit once the energy buffer is full, so it can have an atomically executed energy burst with a minimum energy guarantee. For our head-counting systems, the application runs on the LPC54102 microcontroller. It contains an ARM Cortex-M4F core with 512 kB of flash memory and 104 kB of on-chip SRAM, with no data caches. This poses severe memory constraints for any machine learning application, which must be able to fit its entire model within the 512 kB of flash and all data within the 104 kB of local memory. In both implementations, the system's voltage supply is 2.8 V, the minimum required by the Lepton, and compatible with all other peripherals including the LPC microcontroller, the OV7670 vision sensor, and the BLE radio. This voltage is provided by a buck converter within the EMU. To facilitate interfacing with peripherals, the platform runs at a frequency of 80 MHz. The thermal image acquisition works via the camera's SPI interface, while the visual image is acquired via 8 (parallel) GPIO pins. In both cases, the acquired image is first transferred into SRAM. We implemented the algorithm first described in [16] in Ladybirds C, without machine-dependent optimizations or special instructions. In the case of atomic application execution, the head-detection CNN runs directly on SRAM data without transferring any data to NVM.

## 6 EXPERIMENTAL EVALUATION

### 6.1 Methodology

This section evaluates *Julienning*, our proposed optimization flow for partitioning machine learning applications in batteryless systems. Machine learning applications have energy-intensive processing tasks which can be easily partitioned into energy bursts of arbitrary size. We use *Julienning* to partition the thermal and visual head-counting applications presented in Section 5. To this end, we first characterize the energy and data requirements of the head-counting applications. Then, we evaluate three partitioning schemes, including our optimal *Julienning* method. Lastly, we perform a design-space exploration of both the thermal and visual applications. Using *Julienning*, we effectively sweep a range of energy storage capacities and determine the optimal partitioning that minimizes the total energy.

**Set-up.** All energy characterizations are made using an external DC power supply. The open-source RocketLogger measurement device [40] is used for low-side current measurements of the embedded device. GPIO pins are used to mask power traces and thus determine the energy consumption for individual tasks.

For the partitioning evaluation, the energy costs for the kernels and the external NVM are used, and our optimization tool then calculates the different figures of merit.

**Figures of merit.** For the system performance analysis, the following metrics are used in all experiments:

$N_{bursts}$: The number of bursts necessary to execute the entire application.

$E_s$: The energy required to boot up the system.

$E_r$: The energy required to read (load) the input data for all bursts of a given partitioning.

$E_{kernel,i}$ **(kernel energy):** The energy required to execute the $i$-th kernel/task.

$E_w$: The energy required to write (save) the output data for all bursts of a given partitioning.

$E_{app} = \sum_i E_{kernel,i}$ **(application energy):** The energy required to execute the entire application atomically (without state-retention overheads).

$E_{total} = E_s * N_{bursts} + E_r + E_w + E_{app}$: The total energy required the execute an application with a given partitioning.

$Q_{max}$ **(energy storage bound):** The maximum storage capacity allowed by the system.

$Q_{min}$ **(minimum storage):** The minimum storage capacity needed to reliably execute an application with a given partitioning.

## 6.2  Head-Counting Embedded System

The custom-designed hardware using the thermal- and visual-based cameras can be seen in Figure 5. The only hardware difference between the two versions is the cameras. Consequently, the only difference between head-counting applications executed in both systems is the energy cost of the image acquisition kernel, which is significantly higher for the thermal camera. This difference in energy cost of atomic tasks has a big impact in the partitioning design space. The smaller energy cost of the visual camera allows the entire application to be split into smaller energy bursts. The start-up energy costs, $E_s$, depend only on the microcontroller, which is the same in both platforms. This cost was measured to be 9 $\mu$J.

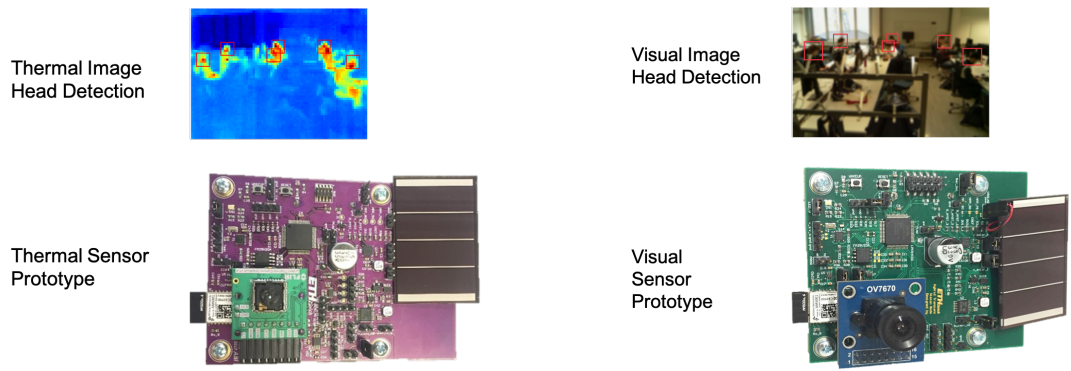

Fig. 5.  Two custom-built batteryless sensing systems. The only difference between the two versions is the camera: one captures temperature, the other visible light.

**Application Memory Requirements.** There are three key data elements: the image itself, the weights of the CNN, and the intermediate results. The Lepton sensor produces an 80 × 60 matrix of 16-bit integers, while the weights and intermediate results used by the CNN are 32-bit floating-point numbers. The code was compiled with the `-Os` optimization flag to minimize its size. The head-detection application (stride 3×3) running on the LPC54102 requires 444 kB, 63 kB, and 186 Bytes for the Text, BSS, and Data sections, respectively. It should be noted that the biggest section, Text, contains all of the constants for the CNN filters, and is stored on the on-chip flash. The BSS and Data sections are small enough to fit comfortably in the available 104 kB SRAM.

**Nonvolatile Memory Costs.** Many nonvolatile memory technologies have asymmetric costs for reading and writing, and thus need to be modeled separately. We model the energy cost for reading and writing to the external FRAM memory linearly with respect to the amount of data, as discussed in Section 4.1. The data transfer on both embedded platforms was optimized for energy efficiency by using direct memory access (DMA). The characterized energy costs of the FRAM memory, $E_r(p)$ and $E_w(p)$, with packet size $|p|$ in Bytes, was modeled as follows: $E_r(p) = 1.3\,\mu\text{J} + |p| \times 7.6\,\text{nJ/Byte}$ and $E_w(p) = 0.9\,\mu\text{J} + |p| \times 6.2\,\text{nJ/Byte}$. This means that saving the entire 80 × 60 thermal picture into FRAM only requires 59.5 $\mu$J.

**Kernel characterizations.** *Julienning* can partition the application to execute one task per burst, thus facilitating energy characterization measurements using a DC source. To characterize the energy requirements for different kernels, a DC source is used to supply 2.8 V. The RocketLogger is connected for low-side measurements and GPIO flags are used to mask the power trace and calculate the kernel energy. For our two sensor prototypes, the only difference is in the image acquisition kernel; all other kernels are the same. Table 1 shows the energy costs of the kernels that use external peripherals. As expected, the thermal camera consumes much more energy than the visual camera. It should be noted that this energy already includes the overhead for turning on the camera. In the case of the Lepton thermal camera, this start-up energy represents over 90% of the total cost.

Table 1. Energy costs for kernels using external peripherals. Measurements were done with $V_{DD}$=2.8 V @ 80 MHz.

| Kernel | Energy per Kernel [mJ] |
|---|---|
| Thermal Image Acquisition | 131.9 |
| Visual Image Acquisition | 4.4 |
| BLE Transmission | 0.086 |

A stride of 3×3 was selected, resulting in a total of ~7300 windows which are processed to determine the total headcount in a full image. The execution of the CNN has a computational complexity of ~50k multiply-accumulate operations for each window. The total execution time for the head counting algorithm is 63 seconds. For more detailed information regarding the CNN's training and architecture, readers are referred to [16]. Table 2 shows the energy necessary to perform the full algorithm including the bulk of CNN computation as well as the pyramid construction, the non-maximum suppression, and the final thresholding. These measurements are based on running the kernels over 100 times. In the case of the CNN kernels, of which there are three types (named CNN1, CNN2, and CNN3) these are executed over 1000 times in total. The non-maximum suppression (NMS) kernel, which collapses overlapping detections into one, is one of smaller steps in the entire pipeline. To determine the necessary capacitor size, the maximum energy consumption for each kernel is used.

Table 2. Energy cost of processing kernels during one complete head-counting application execution (3×3 stride). Measurements were done with $V_{DD}$=2.8 V @ 80 MHz.

| Kernel | $E_{kernel}$ [mJ] | $N_{tasks}$ | $E_{sum}$ [mJ] |
|---|---|---|---|
| Normalize | 0.043 | 1 | 0.043 |
| Initialize | 0.003 | 1 | 0.003 |
| CNN1 | 0.396 | 4125 | 1633.5 |
| CNN2 | 0.396 | 936 | 370.7 |
| CNN3 | 0.403 | 391 | 157.6 |
| Sort | 0.010 | 1 | 0.010 |
| NMS | 0.006 | 1 | 0.006 |
| **Total head-counting** | | | 2161.8 |

## 6.3 Partitioning Results

Once the energy requirements of the individual kernels are known, we can proceed to choose the execution configuration. We evaluate three different algorithms to partition the sequential head-counting application into bursts, comparing *Julienning* to two fixed partitioning schemes: *Single Task* and *Whole Application*. In *Single Task* partitioning, each kernel execution is assigned its own burst and state retention is not optimized, meaning every burst will save and restore all application data. With *Whole Application*, all kernels are assigned to a single burst. Since the image acquisition kernel is the energy-dominant kernel in both versions of the application, the minimum energy storage capacity must be able to guarantee the sensing task (taking the picture) and the image transfer to NVM memory. As shown in the previous subsection, saving the entire image from internal SRAM to external NVM has a negligible cost compared to taking the picture. We thus use $Q_{max}$=132 mJ as the smallest feasible energy capacity.

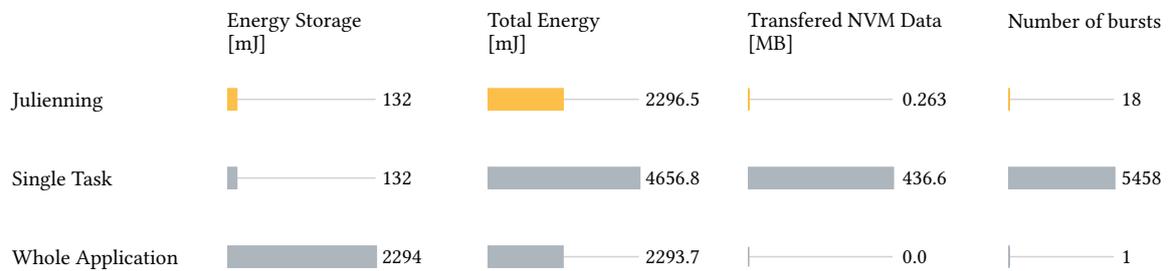

Fig. 6. *Julienning* was evaluated on the thermal application using $Q_{max}$=132 mJ, a fraction of the energy storage required by the *Whole Application* method and enough to guarantee the largest atomic task. With this storage constraint, *Julienning* splits the application into 18 bursts and optimizes the data transfer to NVM, increasing the total energy cost of running the application by only 0.12%.

Figure 6 shows the results for the three partitioning schemes in different figures of merit. *Single Task* partitioning uses the lowest possible storage capacity, but due to its inefficient state-retention scheme, it ends up transferring over 437 MB of data over its 5458 bursts. That causes the energy overhead to be larger than the application energy itself. *Whole Application* minimizes the data overhead since everything is executed within a single burst and no state retention is necessary. However, it requires a large energy storage capacity. Since batteryless sensing systems are designed to minimize storage capacity, the *Whole Application* scheme is not a scalable solution. *Julienning* has the minimum feasible storage capacity, like *Single Task*, and a very low data and burst overhead, similar to *Whole Application*. Since every kernel is specified with explicit data dependencies, every burst loads the data required for its own kernels. Furthermore, our optimization algorithm groups multiple kernels together so that all energy bursts are as close as possible to the minimum storage capacity. In the end, this reduces the number of bursts to just 18, as opposed to 5458 with *Single Task*. Boot-up and data overheads are only 2.79 mJ, or almost 0.1% of the application energy.

**Design-Space Exploration**

To further illustrate the flexibility of our *Julienning* method, we now look at the design exploration of both the visual and thermal head-counting applications. The energy characterization of all individual kernels has been shown in Table 1 and Table 2. We have seen how basic partitioning schemes are task-based, grouping a fixed number of tasks in the same burst. These fixed task schemes have been shown to be inefficient, leading either to large data overhead or very large storage capacities. With *Julienning*, we can select an energy-burst bound ($Q_{max}$), and the algorithm will find the partitioning that minimizes the total application energy within that bound. These values depend on several parameters including the weights of the kernels, the data-transfer costs, and the boot-up costs. In the following, we will see how the $Q_{max}$ parameter impacts the main design metrics, namely $N_{bursts}$ and the energy overhead from partitioning. For each application version, points on the graphs at a given $Q_{max}$ belong to the same solution.

Figure 7 shows the design-space exploration using *Julienning* on the thermal and visual head-counting applications. The x-axis is the $Q_{max}$ used in each optimization run, and the y-axis is the optimal $N_{burst}$. The visual-imaging version exhibits a wider feasibility range due to its fine-grained kernels. This allows *Julienning* to partition it into 456 bursts while incurring an 875.6 mJ overhead. $N_{bursts}$ decreases monotonically with $Q_{max}$, which is expected as increasing energy storage reduces boot-ups and data transfers. Once $Q_{max} > E_{app} + E_{boot-up}$, the optimal $N_{bursts}$ is always 1.

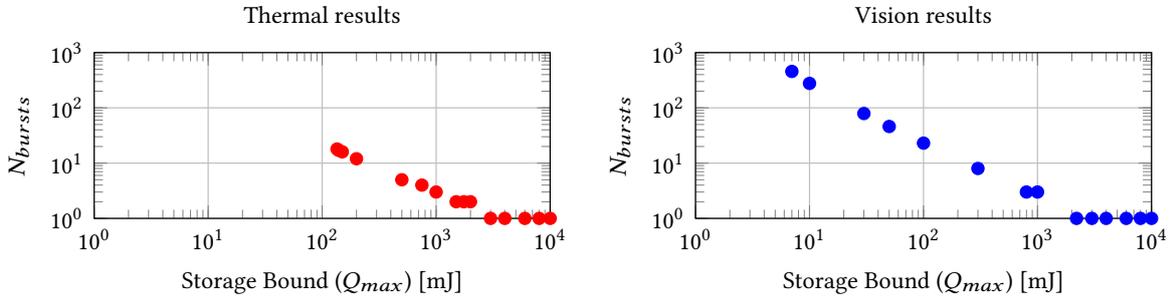

Fig. 7. Log-log plots of *Julienning* partitioning over the feasible design space of both head-counting applications. The visual-imaging version can be partitioned into many more energy bursts because its most energy-intensive atomic task requires only 4.4 mJ, compared to 132 mJ in the thermal-imaging version.

Figure 8 shows the results from *Julienning* when applied to the thermal and visual head-counting applications. The x-axis is the $Q_{max}$ used in each optimization run. Each point represents the optimized partitioning and its resulting $E_{total}$, which includes boot-up, data transfer, and application energy. Again, the visual-imaging version has a wider feasibility range. This is due to the fact that the dominant kernel, the visual image acquisition, requires relatively little energy. As expected, the overhead is smallest when the entire application can be executed in single burst. The energy overhead does increase slowly as the application is partitioned over more bursts. However, *Julienning*'s energy cost optimization keeps the overhead negligibly small for a wide range of storage bounds. As the visual-imaging version gets partitioned into hundreds of bursts, the inevitable overhead of loading and storing data into nonvolatile memory becomes more noticeable.

### 6.4 Analysis of Partitioning Results

Using *Julienning*, we can automate the optimal partitioning of machine learning applications to satisfy a given energy storage limit. The optimization model takes as input the application graph with data dependencies and precharacterized energy costs. Our tool then determines the partitioning that minimizes the total energy while satisfying a storage capacity bound. This easy-to-use optimization flow works for long and complex machine learning applications for which manually finding an optimal solution would not be feasible due to the very large design space.

**Limitations of *Julienning*.** Our proposed methodology has been demonstrated with two versions of a machine learning application running on different hardware prototypes. Both systems have different energy requirements stemming from the characteristics of the camera hardware. The energy cost of the atomic kernels has a direct influence on the granularity with which *Julienning* can partition an entire machine learning application and reduce the required energy storage size. Since kernels are executed atomically, the maximum energy storage reduction is given by the following ratio:

$$\frac{max\left\{E_{load,i} + E_{kernel,i} + E_{save,i}\right\}}{\sum_i E_{kernel,i}} \quad (1)$$

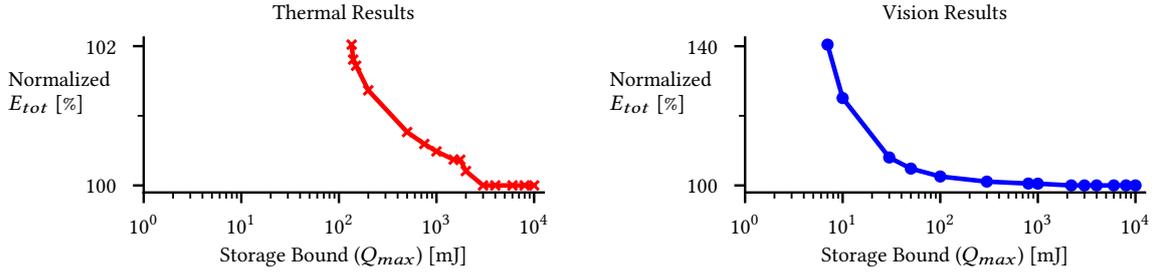

Fig. 8. Semi-log plots of *Julienning* overhead over the feasible design space of both head-counting applications. The overhead introduced by partitioning slowly increases with the number of bursts, staying below 3% for storage bounds as low as 4.3% of $E_{app}$.

The numerator is the largest possible (single-kernel) burst, and the denominator is the total application energy. In the case of the thermal camera, for example, the sensing task is almost 6% of the total energy. For this reason, the thermal application has a smaller feasibility range of 1–18 energy bursts, compared to the 1–456 for the visual application. It should be noted that if the processing tasks were small enough that $E_r + E_w$ dominates burst energy, the benefits of selecting small bursts with few tasks would be diminished. In our system, loading and saving nonvolatile data has a low energy cost compared to the actual processing tasks. This is primarily due to the high energy efficiency and endurance of the selected FRAM chip, but it comes with a high component cost.

**Fine-tuning the energy storage bound.** *Julienning* uses the energy storage bound as a constraint for the optimization. This gives designers the flexibility to choose their energy storage bound according to their system's requirements and environmental conditions. Using *Julienning*, the energy cost will always be minimal, but there will be trade-offs in other metrics. For example, the *Julienning* solution for the thermal-imaging system with $Q_{max}$=132 mJ, discussed in Section 6.3, will have a fast charging time, since the sensing task will execute as soon as 132 mJ have been harvested. If the $Q_{max} = E_{app}$, then the optimal solution will be to run the entire application atomically, and it will have an energy cost slightly below the partitioned version. This atomic solution, however, will perform sensing only after harvesting 2.294 J, which takes significantly more time than the partitioned approach. Designers thus have the possibility to traverse the design space's Pareto front with ease, thanks to *Julienning*'s automated nature.

**Separation of concerns.** The generality of EMU-based designs allowed both machine learning systems to be developed independently from transducer characteristics while still guaranteeing reliable execution. In fact, the only difference between them is the energy storage capacity (i.e., capacitor size). We have seen how different capacity bounds affect important figures of merit like energy overhead, number of bursts, and charge time. Thanks to *Julienning*, arbitrarily long machine learning applications can be executed with a bounded energy storage capacity, in a known number of energy bursts, with optimized data transfers and minimized total energy use.

## 7 SUMMARY

Batteryless systems propose a new paradigm of energy-driven computation that enables long-term deployments of wireless sensors in a scalable, cost-effective, and energy-efficient manner. However, due to the combination of variable primary energy and limited energy storage, batteryless systems must employ a reliable execution model to guarantee that applications execute in a functionally correct manner even when primary energy is sporadic and unreliable. To tolerate these full power failures, batteryless systems must use nonvolatile memory and do so efficiently to minimize overhead. This is particularly challenging in machine learning applications since they require significant memory and computational capacity. In this work, we have developed *Julienning*, a methodology to optimally partition sequential applications executed on energy-harvesting-based systems with a bounded energy storage capacity. To use *Julienning*, software must be specified using a data-flow model with explicit data dependencies. To demonstrate *Julienning*'s power and flexibility, we have implemented two versions of a machine learning application, along with custom-designed hardware, from the computer vision domain. One version of our application estimates the occupancy in a room based on visual images, while the other uses thermal images. Using experimentally characterized energy costs, we transform the partitioning problem into a shortest path problem, easily solvable in polynomial time. Due to explicit data dependencies, *Julienning* can optimize the data transfer for every application partition. This is achieved by loading only the data required for each partition and saving only the data required by future ones. In batteryless systems, where storage capacity needs to be minimized, *Julienning* can be used to rapidly explore the design space of an arbitrarily long sequential application. Using the thermal head-detection application as a benchmark, our proposed methods can reduce the energy storage by 94% compared to no partitioning, while incurring an energy overhead of only 0.12%.

## ACKNOWLEDGMENTS

This work has been supported by the Swiss National Science Foundation, under grant 157048: Transient Computing Systems, as well as the GFF-IPF Grant of the University of St.Gallen's Basic Research Fund.

## REFERENCES

[1] ACKme Networks 2014. *AMS001 AMS002 Datasheet*. ACKme Networks. Preliminary Datasheet.
[2] Mikhail Afanasov, Naveed Anwar Bhatti, et al. 2020. Battery-Less Zero-Maintenance Embedded Sensing at the MithræUm of Circus Maximus. In *Proc. SenSys Conf.* (Virtual Event, Japan). ACM, 368–381.
[3] Domenico Balsamo, Alex S Weddell, Geoff V Merrett, Bashir M Al-hashimi, Davide Brunelli, and Luca Benini. 2015. Hibernus : Sustaining Computation during Intermittent Supply for Energy-Harvesting Systems. *Embed. Syst. Lett. IEEE* 7, 1 (2015). https://doi.org/10.1109/LES.2014.2371494
[4] Naveed Bhatti and Luca Mottola. 2016. Efficient state retention for transiently-powered embedded sensing. In *International Conference on Embedded Wireless Systems and Networks*. 137–148.
[5] Naveed Anwar Bhatti, Muhammad Hamad Alizai, Affan A Syed, and Luca Mottola. 2016. Energy harvesting and wireless transfer in sensor network applications: Concepts and experiences. *ACM Transactions on Sensor Networks (TOSN)* 12, 3 (2016), 24.
[6] Naveed Anwar Bhatti and Luca Mottola. 2017. HarvOS: Efficient Code Instrumentation for Transiently-powered Embedded Sensing. In *Proc. IPSN Conf.* ACM, New York, NY, USA. https://doi.org/10.1145/3055031.3055082
[7] Bernhard Buchli, Felix Sutton, Jan Beutel, and Lothar Thiele. 2014. Towards enabling uninterrupted long-term operation of solar energy harvesting embedded systems. In *European Conference on Wireless Sensor Networks*. Springer, 66–83.
[8] Michael Buettner, Ben Greenstein, and David Wetherall. 2011. Dewdrop: an energy-aware runtime for computational RFID. In *Proc. USENIX NSDI*. 197–210.

[9] Alexei Colin and Brandon Lucia. 2016. Chain: Tasks and Channels for Reliable Intermittent Programs. In *Proc Object-Oriented Programming, Systems, Languages, and Applications Conf.* (Amsterdam, Netherlands). ACM. https://doi.org/10.1145/2983990.2983995

[10] Alexei Colin and Brandon Lucia. 2018. Termination Checking and Task Decomposition for Task-based Intermittent Programs. In *Proc. Compiler Construction Conference* (Vienna, Austria). ACM, New York, NY, USA. https://doi.org/10.1145/3178372.3179525

[11] Alexei Colin, Emily Ruppel, and Brandon Lucia. 2018. A Reconfigurable Energy Storage Architecture for Energy-harvesting Devices. In *Proc. Architectural Support for Programming Languages and Operating Systems Conf.* ACM. https://doi.org/10.1145/3173162.3173210

[12] Cypress Semiconductor Corp. 2015. *1-mbit Serial F-RAM Datasheet.* Cypress Semiconductor Corp. Rev. E.

[13] Edsger W. Dijkstra. 1959. A note on two problems in connexion with graphs. *Numer. Math.* 1, 1 (Dec. 1959), 269–271. https://doi.org/10.1007/bf01386390

[14] FLIR Systems, Inc. 2018. *Lepton Engineering Datasheet.* FLIR Systems, Inc. Rev. 200.

[15] Andres Gomez. 2020. On-demand communication with the batteryless MiroCard: demo abstract. In *Proceedings of the 18th Conference on Embedded Networked Sensor Systems.* 629–630.

[16] Andres Gomez, Francesco Conti, and Luca Benini. 2018. Thermal Image-Based CNN's for Ultra-Low Power People Recognition. In *Proceedings of the Computing Frontiers Conference.* ACM.

[17] Andres Gomez, Christian Pinto, Andrea Bartolini, Davide Rossi, Luca Benini, Hamed Fatemi, and Jose Pineda de Gyvez. 2015. Reducing energy consumption in microcontroller-based platforms with low design margin co-processors. In *Proc. DATE.*

[18] Andres Gomez, Lukas Sigrist, Michele Magno, Luca Benini, and Lothar Thiele. 2016. Dynamic Energy Burst Scaling for Transiently Powered Systems. In *Proc. DATE Conf.* EDA Consortium, 349–354.

[19] Andres Gomez, Lukas Sigrist, Thomas Schalch, Luca Benini, and Lothar Thiele. 2017. Wearable, Energy-Opportunistic Vision Sensing for Walking Speed Estimation. In *Proc. SAS Symp.* IEEE, 1–6.

[20] Andres Gomez, Lukas Sigrist, Thomas Schalch, Luca Benini, and Lothar Thiele. 2018. Efficient, Long-Term Logging of Rich Data Sensors using Transient Sensor Nodes. *ACM Transactions on Embedded Computing Systems* 17, 1 (2018).

[21] Pascal Alexander Hager, Hamed Fatemi, Jose Pineda de Gyvez, and Luca Benini. 2017. A scan-chain based state retention methodology for IoT processors operating on intermittent energy. In *Proc. DATE Conf.* EDA Consortium.

[22] Josiah Hester, Lanny Sitanayah, and Jacob Sorber. 2015. Tragedy of the coulombs: Federating energy storage for tiny, intermittently-powered sensors. In *Proc. Embedded Networked Sensor Systems Conf.* ACM.

[23] Josiah Hester, Kevin Storer, and Jacob Sorber. 2017. Timely Execution on Intermittently Powered Batteryless Sensors. In *Proc. SenSys Conf.*

[24] Jingtong Hu, Chun Jason Xue, Qingfeng Zhuge, Wei-Che Tseng, and Edwin H.-M. Sha. 2013. Write Activity Reduction on Non-volatile Main Memories for Embedded Chip Multiprocessors. *ACM Trans. Embed. Comput. Syst.* 12, 3, Article 77 (April 2013), 27 pages. https://doi.org/10.1145/2442116.2442127

[25] Neal Jackson, Joshua Adkins, and Prabal Dutta. 2019. Capacity over capacitance for reliable energy harvesting sensors. In *Proceedings of the 18th International Conference on Information Processing in Sensor Networks.* 193–204.

[26] Hrishikesh Jayakumar, Arnab Raha, and Vijay Raghunathan. 2014. QUICKRECALL: A Low Overhead HW/SW Approach for Enabling Computations across Power Cycles in Transiently Powered Computers. *Proc. Int. Conf. VLSI Design* (2014). https://doi.org/10.1109/VLSID.2014.63

[27] Sudhanshu Khanna, Steven C Bartling, Michael Clinton, Scott Summerfelt, John A Rodriguez, and Hugh P. McAdams. 2014. An FRAM-Based Nonvolatile Logic MCU SoC Exhibiting 100% Digital State Retention at VDD=0 V Achieving Zero Leakage With <400-ns Wakup Time for ULP Applications. *IEEE J. Solid-State Circuits* 49, 1 (jan 2014). https://doi.org/10.1109/JSSC.2013.2284367

[28] Yongpan Liu, Zhibo Wang, et al. 2016. A 65nm ReRAM-enabled nonvolatile processor with 6x reduction in restore time and 4x higher clock frequency using adaptive data retention and self-write-termination nonvolatile logic. In *Proc. ISSCC Conf.* IEEE. https://doi.org/10.1109/ISSCC.2016.7417918

[29] M Meli and P Bachmann. 2016. *Powering Long Range Wireless Nodes with Harvested Energy.* Technical Report November. Zürcher Hochschule für Angewandte Wissenschaften (ZHAW).

[30] M Meli and U Beerli. 2011. *Indoor battery-less temperature and humidity sensor for Bluetooth Low Energy.* Technical Report November. Zürcher Hochschule für Angewandte Wissenschaften (ZHAW).

[31] Clemens Moser, Lothar Thiele, Davide Brunelli, and Luca Benini. 2010. Adaptive power management for environmentally powered systems. *IEEE Trans. Comput.* 59, 4 (2010), 478–491.

[32] Saman Naderiparizi, Aaron N Parks, Zerina Kapetanovic, Benjamin Ransford, and Joshua R Smith. 2015. WISPCam : A Battery-Free RFID Camera. In *Proc. IEEE RFID.*

[33] NXP Semiconductors 2015. *LPC5410X Product Data Sheet.* NXP Semiconductors. Rev 2.2.

[34] OmniVision 2006. *OV7670/OV7171 CMSO VGA CameraChip Sensor Preliminary Datasheet.* OmniVision. Version 1.4.

[35] Masood Qazi, Ajith Amerasekera, and Anantha P Chandrakasan. 2014. A 3.4-pj feram-enabled d flip-flop in 0.13-$\mu$m CMOS for nonvolatile processing in digital systems. *IEEE J. Solid-State Circuits* 49, 1 (jan 2014), 202–211. https://doi.org/10.1109/JSSC.2013.2282112

[36] Vijay Raghunathan, Aman Kansal, Jason Hsu, Jonathan Friedman, and Mani Srivastava. 2005. Design considerations for solar energy harvesting wireless embedded systems. In *Proceedings of the 4th international symposium on Information processing in sensor networks*. IEEE Press, 64.

[37] Benjamin Ransford, Jacob Sorber, and Kevin Fu. 2011. Mementos: System Support for Long-running Computation on RFID-scale Devices. *SIGPLAN Not.* 46, 3 (March 2011), 12 pages. https://doi.org/10.1145/1961296.1950386

[38] Alberto Rodriguez Arreola, Domenico Balsamo, Geoff V Merrett, and Alex S Weddell. 2018. RESTOP: Retaining External Peripheral State in Intermittently-Powered Sensor Systems. *Sensors* 18, 1 (2018).

[39] Barry K Rosen, Mark N Wegman, and F Kenneth Zadeck. 1988. Global value numbers and redundant computations. In *Proceedings of the 15th ACM SIGPLAN-SIGACT symposium on Principles of programming languages*. ACM, 12–27.

[40] Lukas Sigrist, Andres Gomez, Roman Lim, Stefan Lippuner, Matthias Leubin, and Lothar Thiele. 2017. Measurement and validation of energy harvesting IoT devices. In *Proc. DATE Conf.* IEEE, 1159–1164.

[41] Lukas Sigrist, Andres Gomez, and Lothar Thiele. 2019. Dataset: Tracing Indoor Solar Harvesting. In *Proceedings of the 2nd Workshop on Data Acquisition To Analysis* (New York, NY, USA) *(DATA'19)*. Association for Computing Machinery, New York, NY, USA, 47–50. https://doi.org/10.1145/3359427.3361910

[42] Peter Spies, Markus Pollak, and Loreto Mateu. 2015. *Handbook of energy harvesting power supplies and applications*. CRC Press.

[43] Naomi Stricker and Lothar Thiele. 2020. Analysing and Improving Robustness of Predictive Energy Harvesting Systems. In *Proc. ENSsys Workshop* (Virtual Event, Japan). Association for Computing Machinery, New York, NY, USA.

[44] Texas Instruments 2015. *MSP430FR59xx Mixed-Signal Microcontrollers*. Texas Instruments. Rev. E.

[45] Andreas Tretter. 2018. *On Efficient Data Exchange in Multicore Architectures*. Ph.D. Dissertation. ETH Zurich.

[46] Theodoros D. Verykios, Domenico Balsamo, and Geoff V. Merrett. 2018. Selective policies for efficient state retention in transiently-powered embedded systems: Exploiting properties of NVM technologies. https://doi.org/10.1016/j.suscom.2018.07.003

[47] A. S. Weddell, M. Magno, G. V. Merrett, D. Brunelli, B. M. Al-Hashimi, and L. Benini. 2013. A survey of multi-source energy harvesting systems. In *Proc. DATE Conf.* https://doi.org/10.7873/DATE.2013.190